\documentclass[twocolumn,showpacs,preprintnumbers,amsmath,amssymb]{revtex4}

\usepackage{graphicx}
\usepackage{dcolumn}
\usepackage{bm}
\usepackage{amsmath}

\begin{document}


\title{A TFD model for the Electrospheres of Bare Strange Quark Stars}

\author{Hai-Chuan Guo$^1$, Ren-Xin Xu$^2$, Cheng-Min Zhang$^3$}
\affiliation{$^1$The National Astronomical Observatories, CAS, Beijing 100080; cityseariver@gmail.com\\
$^2$School of Physics, Peking University, Beijing 100871; r.x.xu@pku.edu.cn\\
$^3$The National Astronomical Observatories, CAS, Beijing100080;
zhangcm@bao.ac.cn}

\date{\today}

\begin{abstract}
We study the layer of electrons on bare strange star surfaces,
taking the Dirac exchange-energy into account. Because electrons
are fermions, the electron wave function must be of
exchange-antisymmetry. The Dirac exchange-energy originates,
consequently, from the exchange-antisymmetry of electron wave
functions. This consideration may result in changing the electron
distribution and the electric field on the surface of bare strange
star.
The strong magnetic field effect on the structures of the
electrospheres is also discussed.
\end{abstract}


\maketitle


We are unfortunately not certain about the real state of matter in
pulsar-like stars even 40 years after the discovery of pulsar.
Nuclear matter ({\em neutron stars}) is one of the speculations,
while quark matter ({\em quark stars}) is an
alternative~\citep[e.g.,][]{lp04,weber05,Alcock,Haensel,Glendenning,Cheng,Weber,Cheng
& Harko,Alford & Reddy,Lugones & Horvath}.
The observational features could be very similar for normal neutron
stars and crusted strange quark stars, but the bare quark surface
should be very useful for us to distinguish those two kinds of
compact objects~\cite{Xu07}. It is therefore worth studying the
structure of electrospheres of bare strange quark stars in
details~\citep[for the previous researches, see, e.g.,][]{Usov05}.

If the Witten's conjecture~\cite{Witten} (the strange quark matter
in bulk is the most stable strong interaction system) is correct,
strange quark star with almost equal numbers of $u$, $d$, and $s$
quarks would exist in case that the baryon density is higher than
the nuclear density.
In the strange quark matter, the three flavors of quarks may keep a
chemical equilibrium, and the whole of the quarks will be
electropositive because of massive $s$ quarks. In order to make sure
the strange quark matter is globally electroneutral, there must
exist proper quantity of electrons.~\cite{Xu1}

Alcock et al.~\cite{Alcock} thought that (1) a strange quark star
would accrete inter-stellar matter, and then a crust would form,
wrapping the strange quark star, and (2) a bare strange star can
not manifest itself as a radio pulsar because of being unable to
generate a magnetosphere.
However, it is argued~\cite{Xu2,xu02,cyx07} that there are some
advantages in understanding the observational features if radio
pulsars are bare strange stars and that the circumstellar matter
could not be accreted onto the star's surface at all (i.e., the
quark surface keeps bare).
The electrons can only extend thousands of femto-meters above the
quark surface, and an extremely strong electric field
$E\sim5\times10^{17}$ V/cm forms~\cite{Alcock,Kettner,Xu1,Hu &
Xu}, being higher than the critical field $E_{\rm
cr}\simeq1.3\times10^{16}$V/cm~\cite{Schwinger}.
It is worth noting that the Usov mechanism ($e^{\pm}$-pair
production in degenerate electron gas with strong electric
field)~\cite{Usov1,Usov2,Aksenov} may dominate in the thermal
emission of a bare strange star with a surface temperature $T_{\rm
s}$, $6\times10^8$ K $\lesssim T_{\rm s}\lesssim5\times10^{10}$K.

Sanwal et al.~\cite{Sanwal} discovered two absorption features in
1E 1207-5209's X-ray spectrum, $0.7$ keV, $1.4$ keV. Based on the
works of Ruder et al.~\cite{Ruder} and Mori and Hailey et
al.~\cite{Mori & Hailey}, Sanwal et al. thought the two absorption
features originated from the atomic transitions of once-ionized
helium in the neutron star atmosphere with a strong magnetic
field.
They do not think it is likely to interpret the two absorption
features in terms of electron cyclotron lines for four points of
view.
Howsoever, these points were criticized by Xu et al.~\cite{Xu3},
who suggested that the two absorption features could be of
electron cyclotron lines from the bare strange quark star surface.
Xu et al. proposed a model with a debris disc around a bare
strange quark star, which could slow down the spin and decrease
the inferred strength of the dipole magnetic field so as to fit
the energy differences of Landau levels ($\sim 0.7$ keV).
It is then very necessary for us to understand the exact electron
distribution on the bare strange star surface in terms of
cyclotron lines.

The extreme relativistic electron distribution of the bare strange
star surface has been calculated via Thomas-Fermi
Model~\cite{Alcock,Xu1}, but the Dirac
exchange-energy~\cite{Dreizler} was not taken into account
up-to-now.
The wave function must be of exchange-antisymmetry for fermions,
and the Dirac exchange-energy originates consequently. For the
Dirac exchange-energy plays an important role in the low-density
electrons and make a dramatic effect on the electron distribution,
we need to take it into account, applying the Thomas-Fermi-Dirac
(TFD) model for bare strange stars.

The theoretical basis of the Thomas-Fermi-Dirac Model is the
density functional theory, called DFT for short. Via this theory,
we could obtain the Thomas-Fermi-Dirac Equation of the electron
density.

In DFT, the ground state energy of the system is the functional of
the ground state electron density, which is~\cite{Dreizler}
\begin{equation}
E_0=E_0[n_0(\vec{r})]
\end{equation}
where $E_0$ is the ground state energy of the system, $n_0$ is the
ground state electron density.
The ground state energy of the electron gas is composed of three
parts (we used the different form from ref.~\cite{Dreizler}),
\begin{eqnarray}
E_0[n_0(\vec{r})]&=&-e\int\!n_0(\vec{r})V_0(\vec{r})\textmd{d}\vec{r}\nonumber\\
& &+\langle\Psi_0|\frac{1}{2}\sum_{i\neq{j}}\frac{e^2}{|\vec{r_i}-\vec{r_j}|}|\Psi_0\rangle\nonumber\\
& &+\langle\Psi_0|\sum_{i}\hat{T}_i|\Psi_0\rangle,
\end{eqnarray}
where $|\Psi_0\rangle$is the ground state vector of the electron
gas, the first term of right side is the outer potential energy of
the system, the second term is the inter-electronic Coulomb energy
and the last one is the electrons kinetic energy.
Here we can divide the inter-electronic Coulomb energy into two
terms, the classical inter-electronic Coulomb energy
\begin{equation}
\frac{e^2}{2}\int\!\frac{n_0(\vec{r})n_0(\vec{r}')}{|\vec{r}-\vec{r}'|}\textmd{d}\vec{r}\textmd{d}\vec{r}'\nonumber,
\end{equation}
and the inter-electronic exchange-energy, also called Dirac
exchange-energy, $E_{\rm ex}$.
And then we can put the terms of the outer potential energy and
the classical inter-electronic Coulomb energy together
\begin{equation}
-e\int\!n_0(\vec{r})V(\vec{r})\textmd{d}\vec{r}\nonumber,
\end{equation}
while putting the terms of the Dirac exchange-energy and the
electron kinetic energy together
\begin{equation}
\int\!G[n_0(\vec{r})]\textmd{d}\vec{r}\nonumber,
\end{equation}
where $G[n_0(\vec{r})]$ is also the functional of the ground state
electron density $n_0(\vec{r})$.
So we get the expression
\begin{eqnarray}
E_0[n_0(\vec{r})]&=&-e\int\!n_0(\vec{r})V(\vec{r})\textmd{d}\vec{r}\nonumber\\
& &+\int G[n_0(\vec{r})]\textmd{d}\vec{r}.
\end{eqnarray}
For the ground state energy is the minimum of the system, it must
be content with the variational equation~\cite{Dreizler} {\footnotesize
\begin{eqnarray}
\delta E_0[n_0(\vec{r})]=-e\int \delta n_0(\vec{r})V(\vec{r})\textmd{d}\vec{r}+\int \delta G[n_0(\vec{r})]\textmd{d}\vec{r}\nonumber\\
=-e\int \delta n_0(\vec{r})V(\vec{r})\textmd{d}\vec{r}+\int
\frac{\delta G[n_0(\vec{r})]}{\delta n_0(\vec{r})}\delta
n_0(\vec{r})\textmd{d}\vec{r}=0.
\end{eqnarray} }
As the electron number is conservative, it must be content with
the variation equation~\cite{Dreizler}
\begin{eqnarray}
\delta N=\int \delta n_0(\vec{r})\textmd{d}\vec{r}=0,
\end{eqnarray}
where $N$ is the total number of electron.
If we take a Lagrange
multiplier~\cite{Dreizler}, $-\mu$, called chemical potential, we can joint the
two variational 0-equations above into one variational
differential equation,
\begin{eqnarray}
-eV(\vec{r})+\frac{\delta G[n_0(\vec{r})]}{\delta
n_0(\vec{r})}=\mu.
\end{eqnarray}

In classical electrodynamics, the Poisson equation reads,
\begin{equation}
\nabla^2V=-4\pi\rho_e,
\end{equation}
where $V$ is the potential produced by the electrons and the
quarks, while $\rho_e$ is the charge density of the electrons and
the quarks. However, out of the strange quark matter surface,
there are no quarks.
If we take the variational differential
equation into the Poisson equation, we'll obtain the
Thomas-Fermi-Dirac equation,
\begin{equation}
\nabla^2\frac{\delta G[n_0(\vec{r})]}{\delta n_0(\vec{r})}=4\pi
e^2(n_0(\vec{r})-n_q),
\end{equation}
where $e(n_o(\vec{r})-n_q)=-\rho_e$, $en_q$ is the charge density
of quark.
For simplicity, we would assume that the strange quark matter is
uniform global electropositive, and $n_q$ is then a constant
$n_q=n_{qp}$ in the strange quark matter, while out of the strange
quark matter surface, $n_q=0$.
For the thickness of the electron gas out of the
quark matter surface is so thin to the radius of the strange quark
star that we can ignore the curvature of the star surface, we can
take $\frac{\textmd{d}}{\textmd{d}z}$ instead of $\nabla$.
If we
are able to obtain the apparent relation between $G[n_0(\vec{r})]$
and $n_0(\vec{r})$, we can take $\textmd{d}$ instead of $\delta$.
So we get the new-formed TFD equation
\begin{equation}
\left(\frac{\textmd{d}}{\textmd{d}z}\right)^2\frac{\textmd{d}G(n_0)}{\textmd{d}n_0}=4\pi
e^2(n_0-n_q),
\end{equation}
where $z$ is the height to the quark matter surface. When $z>0$,
it is out of the surface, $z<0$, it in the strange quark matter,
$z=0$, it on the surface.
If the both sides of the equation above are multiplied by
$\textmd{d}(\frac{\textmd{d}G(n_0)}{\textmd{d}n_0})$(see
ref.~\cite{Dreizler}), it can be transformed into a 1st-order
ordinary differential equation{\scriptsize
\begin{equation}
\textmd{d}\left(\frac{1}{2}\left(\frac{\textmd{d}}{\textmd{d}z}\frac{\textmd{d}G(n_0)}{\textmd{d}n_0}\right)^2\right)=\textmd{d}\left(4\pi
e^2\left(\left(n_0-n_q\right)\frac{\textmd{d}G(n_0)}{\textmd{d}n_0}-G(n_0)\right)\right).
\end{equation}}
It can be inferred that at the infinite height, $n_0=0$ and
${\textmd{d}n_0}/{\textmd{d}z}=0$, so out of the strange quark
star surface, the equation above can be integrated from the
infinite height, simplified to the expression (for $z>0$)
\begin{equation}
\frac{1}{2}\left(\frac{\textmd{d}}{\textmd{d}z}\frac{\textmd{d}G(n_0)}{\textmd{d}n_0}\right)^2=4\pi
e^2\left(n_0\frac{\textmd{d}G(n_0)}{\textmd{d}n_0}-G(n_0)\right).
\end{equation}

In the equation above, the left side of the equal sign is a
square, in order to make sure the equation has a real number
solution, the right side must be lager than or equal to zero,
i.e.,
\begin{equation}
n_0\frac{\textmd{d}G(n_0)}{\textmd{d}n_0}-G(n_0)\geqslant0.
\end{equation}
If we obtain the exact relation between $G(n_0)$ and $n_0$, we can
calculate the minimum of $n_0$ via the in-equation above.
With the increase of the height, $n_0$ decreases continuously to
the point of its minimum and, after reaching the point, will jump
to zero directly for appropriating to the boundary condition.

Deep in the strange quark matter, $z\rightarrow -\infty$, it can
be inferred that the electron charge density equals to the three
flavors of quarks charge density~\cite{Xu1},
\begin{equation}
n_0(z=-\infty)=n_q.
\end{equation}
If we integrate the equ.$(17)$ from $z=-\infty$,we 'll get
{\footnotesize
\begin{equation}
\frac{1}{2}\left(\frac{\textmd{d}}{\textmd{d}z}\frac{\textmd{d}G(n_0)}{\textmd{d}n_0}\right)^2=4\pi
e^2\left(n_0-n_{qp}\frac{\textmd{d}G(n_0)}{\textmd{d}n_0}-\left(G(n_0)-G(n_{qp})\right)\right).
\end{equation}}
On the strange quark matter surface, $z=0$, the electron density
and the gradient of the electron density are continuous, so from
the equ.$(11)$ and equ.$(14)$, we get the equation below
{\scriptsize\begin{eqnarray}
n_0\frac{\textmd{d}G(n_0)}{\textmd{d}n_0}-G(n_0)&=&(n_0-n_{qp})\frac{\textmd{d}G(n_0)}{\textmd{d}n_0}-\left(G(n_0)-G(n_{qp})\right),\\
0&=&-n_{qp}\frac{\textmd{d}G(n_0)}{\textmd{d}n_0}+G(n_{qp}),\\
\left.\frac{\textmd{d}G(n_0)}{\textmd{d}n_0}\right|_{z=0}&=&\frac{G(n_{qp})}{n_{qp}}.
\end{eqnarray}}
Consequently, if we obtain the expression of $G(n_0)$ and the
value of $n_{qp}$, we could work out $n_0$ via the equ.$(17)$ as
the boundary condition of the equ.$(11)$.
Howsoever, we describe the
strange quark matter by its chemical potential that is also called
Fermi energy,
\begin{equation}
\mu_q=\hbar c(3\pi^2n_{qp})^{\frac{1}{3}},
\end{equation}
so we could work out $n_{qp}$ from $\mu_q$.

Usually, the local density approach (LDA) is an effective way to
describe the Dirac exchange-energy.
In this way, the Dirac exchange-energy reads~\cite{Kohn}
\begin{equation}
E_{\rm ex}(n_0)=-2e^2\left(\frac{3n_0}{8\pi}\right)^{\frac{1}{3}}.
\end{equation}

Note that $G(n_0)$ is composed of the Dirac exchange-energy
$E_{\rm ex}(n_0)$ and the electron kinetic energy $T(n_0)$.
In the extreme relativistic case for electron, we could obtain the
electrons kinetic energy via Fermi-Dirac statistics
\begin{equation}
\langle T\rangle=\frac{3}{4}\hbar c(3\pi^2n_0)^{\frac{1}{3}}.
\end{equation}

Actually, we've ignored the influence caused by temperature. While
for the generally relativistic case, the means of the electrons
kinetic energy could also be worked out via Fermi-Dirac statistics
{\scriptsize
\begin{eqnarray}
\langle T\rangle=\Bigg(\left(\frac{3}{4}+\frac{3}{8}(3\pi^2n_0)^{-\frac{2}{3}}\left(\frac{\hbar}{m_ec}\right)^{-2}\right)\sqrt{1+(3\pi^2n_0)^{\frac{2}{3}}\left(\frac{\hbar}{m_ec}\right)^2}\nonumber\\
-\frac{3}{8}(3\pi^2n_0)^{-1}\left(\frac{\hbar}{m_ec}\right)^{-3}\sinh^{-1}\left((3\pi^2n_0)^{\frac{1}{3}}\left(\frac{\hbar}{m_ec}\right)\right)-1\Bigg)m_ec^2.
\end{eqnarray}
}

From the definition of $G(n_0)$, we get the exact relation among
$G(n_0)$, $E_{xc}(n_0)$ and $\langle T\rangle(n_0)$
\begin{equation}
G(n_0)=n_0\{(E_{xc}(n_0)+\langle T\rangle(n_0)\}.
\end{equation}

For the extreme relativistic case, with the expression of
$G(n_0)$, TFD differential equation and the typical chemical
potential (e.g., $20$ MeV~\cite{Xu1}) of the quark in the strange
quark matter, we could calculate the electron distribution out of
the strange quark matter surface.
If we ignore the Dirac
exchange-energy, TFD model will degenerate to TF model calculating
the extreme relativistic electron distribution out of the bare
strange star surface.
For the generally relativistic case, with the similar conditions,
we could work out the distribution.
The distributions of the two models TF and TFD of the two quark
chemical potentials, $2$ MeV and $20$ MeV, are illustrated in the
Fig.1. Also, we can work out the electric field, illustrated in
the Fig.2.
The maximum of it is $\sim10^{17}$V/cm. If we enlarge the range of
the typical chemical potential of the quark in the strange quark
matter, choosing $2$ MeV, $20$ MeV and $200$ MeV separately, we
could get three kinds of the distribution, illustrated in the
Fig.3.

As in Fig.1, in the high-density domain,
TFD-generally-relativistic electron density is about 2 orders of
magnitude more than TF-extremely-relativistic electron density,
while TFD-electron density is much less than TF-electron density
in the low-density domain.
The differences reflect the influence of Dirac exchange-energy. It
is the Dirac exchange-energy that makes the electrons high above
much lower, and it seems to put an attractive force. In Fig.3, we
observe that, in the low-density domain, the electron
distributions of different quark chemical potentials (i.e.,
$V_{\rm q}$) have few differences.

If the strong magnetic field is taken into account, the sum of the
electron kinetic energy and magnetic energy is quantized~\cite{Jaikumar},
\begin{equation}
E=\sqrt{m_e^2c^4+c^2p_z^2+m_e^2\hbar \omega_B\tilde{n}},
\end{equation}
here $\tilde{n}=n_L+\sigma+\frac{1}{2}$, $n_L$ is the orbit
quantum number, $\sigma=\pm 1/2$ is the 3rd component quantum
number of the electron spin, and $\omega_B=eB/mc$ is the cyclotron
angular frequency in the magnetic field.
With the increase of the height, the electron density $n(z)$ will
decrease to a critical density of $n^*$~\cite{Jaikumar},
\begin{equation}
n^*=\frac{2m_e^2\omega_Bc}{h^2}\sqrt{\frac{2\hbar\omega_B}{m_ec^2}}.
\end{equation}
The electrons above the height of $n^*$ will play an important
part in the abstraction of X-ray~\cite{Jaikumar}.
If we take the standard strength $10^{12}$G of the dipole magnetic
field, we can work out $n^*\sim10^{-10}$fm$^{-3}$.
In Fig.3, we can find out the point of $n^*/N=10^{-2}$. Only the
electrons above the height of $n^*$ can absorb the X-rays emitted
from the strange quark matter.
Because TFD-electron density decreases more rapidly after the
$n^*$ than TF-electron density, there are much fewer electrons
which can absorb the X-rays. Therefore, one has to take the Dirac
exchange-energy into account when one tries to study the electron
distribution and the X-ray absorbtion features.

In Fig.3, the point of $n^*/N=10^{-2}$ for $B\sim 10^{12}$ G could
be found and the three kinds of the electron distribution in
low-density domain are similar, the distribution of the electrons,
which are able to absorb the X-rays, is then not sensitive to the
chemical potential of the strange quark matter ($V_{\rm q}$, for
instance).
Nevertheless, if the $n^*$ were larger, the distribution would
become sensitive to the chemical potential of the strange matter.
It is then possible that we might obtain the strange quark matter
chemical potential by studying the X-ray absorbtion features.
In equ.(24),
$n^*=\frac{1}{\sqrt{2}\pi^2}\left(\frac{eB}{c\hbar}\right)^{\frac{3}{2}}\propto
B^{\frac{3}{2}}$, we may need a strong dipole magnetic field,
$B>10^{13}$ G ($n^*/N=10^{-1}$).
In one word, if we observe the bare strange star with the stronger
magnetic field, the absorption features of its X-ray spectrum may
provide us a possibility to know its inner strange matter chemical
potential.

Summarily, the issues studied in the paper make it necessary for
us to take the Dirac exchange-energy into account when we consider
the electrosphere of bare strange stars. The differences between
the electron densities in TF and TFD models are significant. If
the exact distribution of the electrons on the bare strange star
surface is obtained and if the magnetic field of the bare strange
star is strong ($B>10^{13}$ G), one may probably obtain the
electron chemical potential of the strange quark matter by
investigating the thermal absorption spectra.

The authors thank the members in the pulsar group of Peking
University for helpful discussions. This work is supported by NSFC
(10573002, 10778611), the Key Grant Project of Chinese Ministry of
Education (305001), the National foundation for Fostering Talents
of Basic Science (NFFTBS), and the program of the Light in China's
Western Region (LCWR, No. LHXZ200602).


\begin{figure}
\resizebox{7cm}{7cm}{\includegraphics{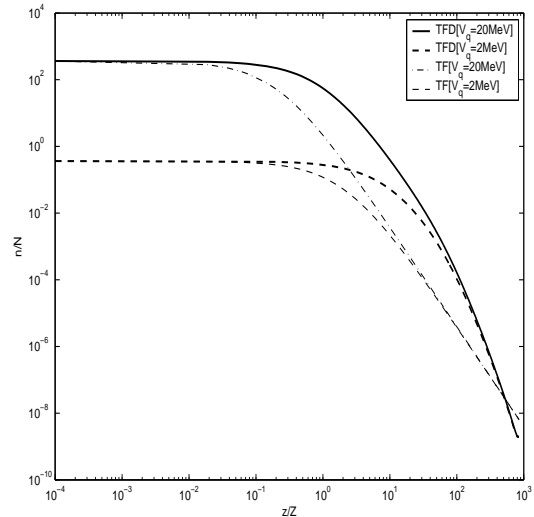}}
\caption{\label{fig:split} This figure illustrates the differences
of the electron density among the different models with different
quark chemical potentials.
$N=\left(\frac{m_ec}{\hbar}\right)^3=1.7\times10^{-8}$fm$^{-3}$.
$Z=\frac{\hbar}{m_ec}=3.9\times10^2$fm. $z$ is the height to the
strange quark matter surface.
$n$ is the electron number density of $z$. In the extremely
relativistic case, the difference between the electron density
curves of TF and TFD models is not significant when the density
$n$ is very high, but is significant when $n$ is lower.
In the high-density domain, TFD-generally-relativistic electron
density is about 2 orders of magnitude more than
TF-extremely-relativistic electron density, while TFD-electron
density is much less than TF-electron density in the low-density
domain.}
\end{figure}

\begin{figure}
\resizebox{7cm}{7cm}{\includegraphics{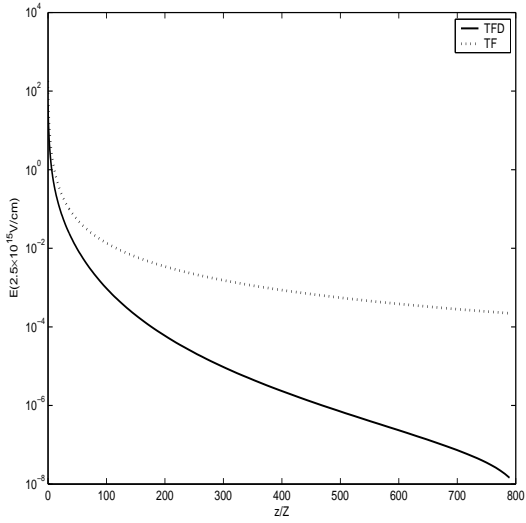}}

\caption{\label{fig:split} This figure illustrates the differences
between the electric fields, $E$, in the TFD model and the TF
model.
The maximums of the electric field in the two models are both
stronger than the critical field $\sim 10^{16}$V/cm.
$Z=\frac{\hbar}{m_ec}=3.9\times10^2$fm.}

\end{figure}

\begin{figure}
\resizebox{7cm}{7cm}{\includegraphics{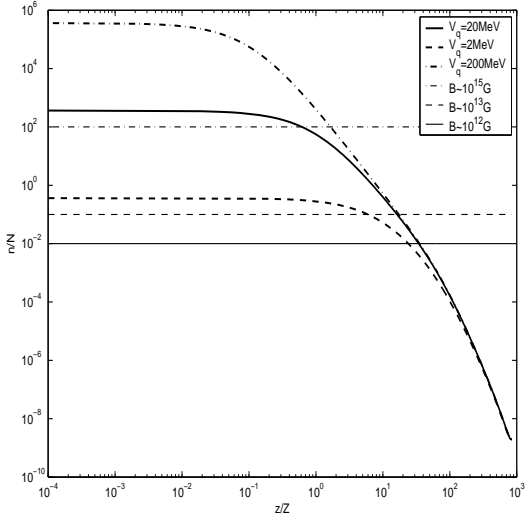}}
\caption{\label{pic:rSplit} This figure illustrates the
differences of the electron density among the different quark
chemical potential in the TFD model.
$z$ is the height to the strange quark matter
surface. $n$ is the electron number density of $z$.
In the high-density domain, the electron distributions of
different quark chemical potentials are very different.
Nevertheless, in the low-density domain, the three electron
distributions have few differences.
The lowest horizontal line is correspondent to
$n^*(B=10^{12}$G$)$.
The middle horizontal line, which is to $n^*(B=10^{13}$G$)$, could
tell the difference between $V_q=20$ MeV and $V_q=2$ MeV. The
highest horizontal line, which is to $n^*(B=10^{15}$G$)$, may tell
the difference between $V_q=200$ MeV and $V_q=20$ MeV.}
\end{figure}

\end{document}